\documentclass[aps,prl,reprint]{revtex4-2}
\usepackage{blindtext}
\usepackage{graphicx}
\usepackage{subcaption}
\usepackage{xcolor}
\usepackage[fleqn]{amsmath}
\usepackage[font=small,skip=3pt]{caption}
\usepackage{makecell}
\usepackage{hyperref} 
\usepackage{cleveref}
\usepackage{url}
\begin{document}
\title{Kinetic-Ballooning-Limited Pedestals in Spherical Tokamak Plasmas}

\author{J. F. Parisi$^{1}$}
\email{jparisi@pppl.gov}
\author{W. Guttenfelder$^{1,2}$}
\author{A. O. Nelson$^3$} 
\author{R. Gaur$^4$}
\author{A. Kleiner$^1$} 
\author{M. Lampert$^1$}
\author{G. Avdeeva$^{5}$}
\author{J. W. Berkery$^1$}
\author{C. Clauser$^6$} 
\author{M. Curie$^4$}
\author{A. Diallo$^1$}
\author{W. Dorland$^7$}
\author{S. M. Kaye$^{1}$}
\author{J. McClenaghan$^{5}$}
\author{F. I. Parra$^{1}$}
\affiliation{$^1$Princeton Plasma Physics Laboratory, Princeton University, Princeton, NJ, USA}
\affiliation{$^2$\textcolor{black}{Type One Energy, 8383 Greenway Boulevard, Middleton, WI, USA} }
\affiliation{$^3$Department of Applied Physics and Applied Mathematics, Columbia University, New York, NY, USA}
\affiliation{$^4$Department of Mechanical and Aerospace Engineering, Princeton University, Princeton, NJ, USA}
\affiliation{$^5$General Atomics, P.O. Box 85608, San Diego, CA, USA}
\affiliation{$^6$Plasma Science and Fusion Center, Massachusetts Institute of Technology, Cambridge, MA, USA}
\affiliation{$^7$Department of Physics, University of Maryland, College Park, MD, USA}

\begin{abstract}
A theoretical model is presented that for the first time matches experimental measurements of the pedestal width-height Diallo scaling in the low-aspect-ratio high-$\beta$ tokamak NSTX. Combining linear gyrokinetics with self-consistent pedestal equilibrium variation, kinetic-ballooning, rather than ideal-ballooning plasma instability, is shown to limit achievable confinement in spherical tokamak pedestals. Simulations are used to find the novel Gyrokinetic Critical Pedestal constraint, which determines the steepest pressure profile a pedestal can sustain subject to gyrokinetic instability. Gyrokinetic width-height scaling expressions for NSTX pedestals with varying density and temperature profiles are obtained. These scalings for spherical tokamaks depart significantly from that of conventional aspect ratio tokamaks.
\end{abstract}

\maketitle

\setlength{\parskip}{0mm}
\setlength{\textfloatsep}{5pt}

\setlength{\belowdisplayskip}{6pt} \setlength{\belowdisplayshortskip}{6pt}
\setlength{\abovedisplayskip}{6pt} \setlength{\abovedisplayshortskip}{6pt}

\textit{Introduction.}-- Fusion energy is a grand challenge of physics and engineering. The tokamak, a prime candidate for magnetic confinement fusion, achieved its fusion power records \cite{Keilhacker1999,Maggi2023} operating in H-mode, a high-confinement regime characterized by much steeper pressure gradients in the pedestal at the plasma edge \cite{Wagner1982,Kaye1984,Ryter1996}. Since fusion power scales approximately quadratically with the pedestal pressure height, accurate prediction of pedestal structure is crucial for upcoming burning plasma experiments such as ITER and SPARC \cite{Shimada2007,Wenninger2015,Sorbom2015,Hughes2020,Rodriguez-Fernandez2022,Osborne2023}. In this work, we make significant advances in pedestal predictive capabilities, allowing accurate predictions across a much wider range of tokamak operating space. In the low-aspect-ratio ($A = R/a$ where $R$ is the plasma major radius and $a$ is the plasma half-diameter) tokamak NSTX, edge pedestals are much wider than in higher aspect-ratio plasmas. The new work presented here yields the first understanding of this difference. These results have wide-ranging implications for the physics basis, design, and optimization of fusion pilot plants, and motivates further investigation into kinetic effects on ballooning modes.

In this Letter, we present the first theoretical model for the pedestal width $\Delta_{\mathrm{ped}}$ and normalized height $\beta_{\theta,\mathrm{ped}}$ to agree with experimental measurements for the National Spherical Torus Experiment (NSTX) \cite{Diallo2013}. The Diallo scaling for NSTX, $\Delta_{\mathrm{ped}} \simeq 0.4 \beta_{\theta,\mathrm{ped}}^{1.05}$, gives pedestals that are much wider than conventional-aspect-ratio tokamaks, $\Delta_{\mathrm{ped}} \simeq 0.08 \sqrt{ \beta_{\theta,\mathrm{ped}}}$ \cite{Snyder2009}. We calculate gyrokinetic stability in spherical tokamak (ST)  \cite{Peng1986, Gryaznevich1998, Menard2016} plasmas with self-consistent equilibrium variations to answer the questions: (1) what additional physics is required for pedestal width-height prediction in STs? (2) what are the steepest-possible pedestal pressure profiles subject to kinetic instability? (3) what are the resulting pedestal width-height scalings?

While H-mode has a higher power density, it often has unstable edge-localized-modes (ELMs) \cite{Kirk2004} and therefore unacceptably high divertor fluxes, necessitating robust ELM-free pedestal regimes \cite{Osborne2015,Chen2017}. Kinetic-ballooning-limited pedestals might present such a regime. We use a gyrokinetic model to predict pedestal width-height scalings, showing that low-aspect-ratio high-normalized-pressure ($\beta$) plasmas in NSTX pedestals are limited by kinetic rather than ideal-ballooning stability. Starting from experimental equilibria, we construct and analyze equilibria with rescaled pedestal width and height. We use gyrokinetic simulations to obtain a profile constraint that matches NSTX width-height measurements.

\textit{Pedestal Models and Regimes.}-- The EPED model \cite{Snyder2002,Snyder2009,Snyder2011} posits that an ELMy pedestal's pressure width $\Delta_{\mathrm{ped}}$ and height $\beta_{\theta,\mathrm{ped}}$ are determined by two stability thresholds; \textcolor{black}{nearly local kinetic-ballooning-modes (KBMs) \cite{Tang1980,Snyder2009}, approximated by} infinite toroidal mode number ($\infty-n$) ballooning \textcolor{black}{stability} \cite{Connor1979,Connor1998} -- limiting pressure gradient -- and peeling-ballooning modes \cite{Lortz1975,Connor1998,Wilson1999} -- limiting current and pressure gradient -- are often in good agreement with experiment \cite{Snyder2009,Snyder2011,Beurskens2011,Walk2012,Leyland2013}. Recent models include additional transport physics such as core transport and EPED coupling \cite{Saarelma2017,Meneghini2021} and stiff edge electron transport \cite{Luda2020,Field2023}. Some pedestal regimes deviate from ideal-MHD predictions, however \cite{Horvath2018,Smith2022}. In low-aspect-ratio, high-$\beta$ NSTX plasmas, measured pedestals \cite{Diallo2013} are much wider than EPED's constraint \cite{Snyder2009} indicating the presence of additional transport mechanisms or other physics. \textcolor{black}{Database analysis of another low-aspect-ratio device, MAST, found ELMy pedestals roughly double the width predicted by the conventional-aspect-ratio scaling, although there was considerable uncertainty in the width-height scaling fit parameters \cite{Smith2022}.  Earlier MAST analysis found $\Delta_{\mathrm{ped}} \sim \sqrt{\beta_{\theta,\mathrm{ped}}}$ for ELMy H-modes with much lower fitting uncertainty, albeit still with much wider pedestals than for conventional-aspect-ratio \cite{Kirk2009}.} Recent work has shown non-ideal effects to be important for peeling-ballooning stability in NSTX \cite{Kleiner2021,Kleiner2022} and for wide pedestals in DIII-D \cite{Chen2017}. Additionally, while ideal-MHD models constrain pressure, they do not specify the relative density and temperature contributions required to predict kinetic instabilities and transport
\cite{Guttenfelder2021}.

\textit{Gyrokinetics and ideal-ballooning.}-- We study plasmas using the gyrokinetic codes GS2 and CGYRO \cite{Dorland2000,Candy2016,Barnes2021}, evolving the distribution function $h_s = q_s \phi ^{tb} F_{Ms} /T_s   + f _s^{tb}$ and the electrostatic and gyrokinetic potentials $\phi ^{tb}$ and $ \chi_s ^{tb}$ \cite{Abel2013} according to the linear gyrokinetic equation,
\begin{equation}
  \frac{  \omega h_{s} + i \mathbf{v}_{\parallel} \cdot \nabla h_s + i \mathbf{v}_{Ms} \cdot \nabla h_s - i C_s^l } { \omega - \omega_{*s}} \\ = \frac{ q_s \chi_s ^{tb} F_{Ms}}{ T_s} ,
    \label{eq:gke}
\end{equation}
and Maxwell's equations \cite{Catto1978,Frieman1982,Parra2008,Abel2013}. Here, a species $s$ has a total turbulent distribution function $f ^{tb} _s$ and a Maxwellian $F_{Ms}$ with temperature $T_s$, $\omega$ is the frequency, $\omega_{*s}$ contains equilibrium instability drives \cite{Hardman2022}, $\Omega_s = q_s B / m_s c $ is the gyrofrequency, $q_s$ is the electric charge, $m_s$ is the mass, $\mathbf{v}_{\parallel}$ and $\mathbf{ v}_{Ms}$ are the parallel and magnetic drift velocities, and $C^{l}_s$ is a pitch angle scattering and energy diffusion collision operator. We evolve all three electromagnetic fields.

Existing pedestal models \cite{Snyder2009,Saarelma2017} give width-height predictions using the ideal-ballooning equation \cite{Connor1979},
\begin{equation}
    d/d \theta [ g \; d X/d \theta ] + d X =  - \omega^2 \lambda X,
    \label{eq:ideal}
\end{equation}
for local stability, which is the ideal limit of \Cref{eq:gke} \cite{Tang1980}. Here, $X$ is the eigenfunction and $g$, $d$, and $\lambda$ are geometric coefficients \cite{Gaur2023}. In this work, we solve \Cref{eq:ideal} using the BALOO \cite{Miller1997} and ball\_stab \cite{Barnes2021,Gaur2023a} codes, which are in excellent agreement, finding that the ballooning stability thresholds of \Cref{eq:gke,eq:ideal} differ significantly for NSTX. The lower kinetic stability threshold (under certain conditions) has been discussed extensively before \cite{Tang1980, Hastie1981, Dong1999, Pueschel2008, Ma2017, Aleynikova2017}. %

\begin{figure}
    \centering
    \includegraphics[width=0.5 \textwidth]{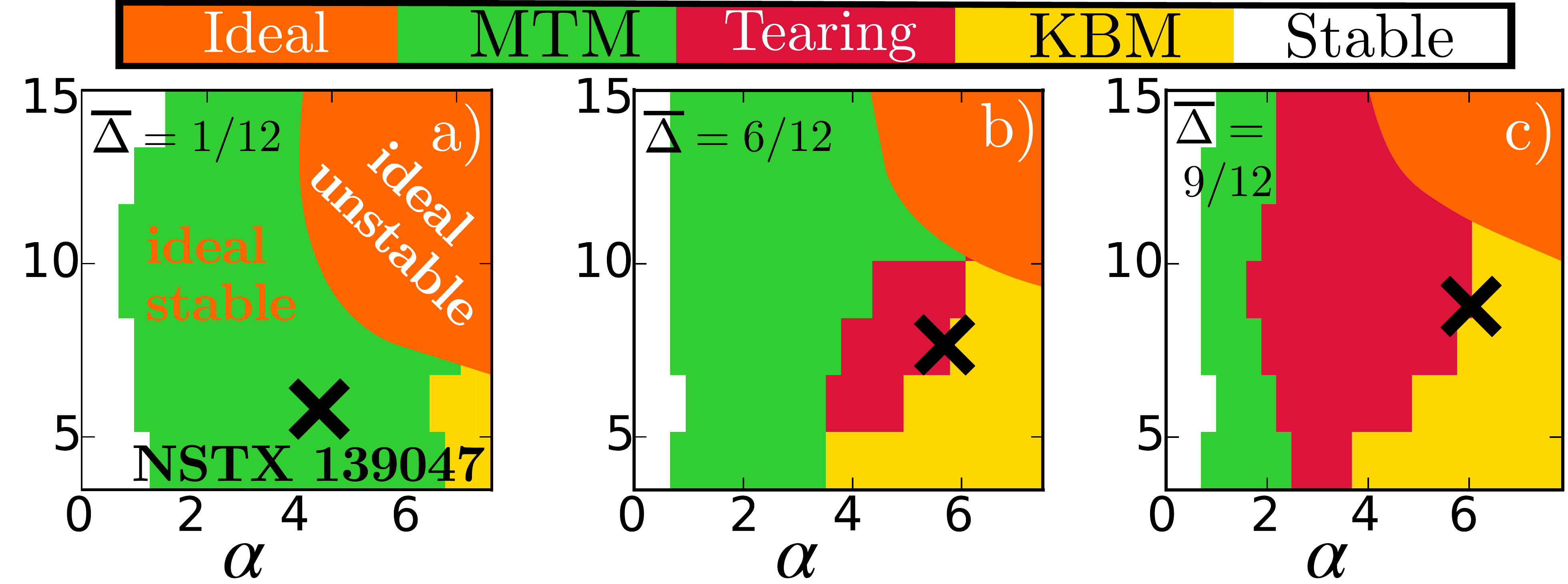}
    \caption{Gyrokinetic and ideal-ballooning $s-\alpha$ analysis for experimental NSTX discharge 139047: the pedestal half-width $\overline{\Delta} \in [3/12, \ldots, 9/12] $ is marginally KBM stable for $k_y \rho_i = 0.18$. X marks experimental points.}
    \label{fig:one}
\end{figure}

\textit{Kinetic-Limited Experiments.}-- We begin our analysis with ELMy NSTX discharge 139047, showing that its pedestal profiles are kinetic-limited (rather than ideal-limited) using local $s-\alpha$ gyrokinetic and ideal-ballooning analysis at 3 radial pedestal locations. $s-\alpha$ analysis perturbs the magnetic shear $s = (r/q)(dq/dr)$ and normalized pressure gradient \textcolor{black}{$\alpha = (2/\pi c^2)(\partial V / \partial \psi) \sqrt{V/(2\pi^2 R_0)} (dp/d\psi)$} on a flux surface while varying the local geometry self-consistently \cite{Greene1981}. \textcolor{black}{We emphasize that this $s-\alpha$ analysis uses realistic geometry and not the shifted-circle approximation in \cite{Connor1978}.} Here, $r$ is the minor radial coordinate, $q$ is the safety factor, \textcolor{black}{$c$ is the speed of light, $V$ is the volume, $R_0$ is the major radius, and $p$ is the pressure.} For the $s-\alpha$ scan in \Cref{fig:one}, we vary temperature and density gradients by the same factor. Shown in \Cref{fig:one}(a)-(c), in order of increasing pedestal radial fraction,
\begin{equation}
\overline{\Delta} = (\psi_N - \psi_{\mathrm{ped}})/\Delta_{\mathrm{ped}},
\end{equation}
we plot the fastest growing gyrokinetic mode for binormal wavenumber $k_y = 0.18/\rho_i$ where the deuterium gyroradius is $\rho_i = \sqrt{T_e / m_i \Omega_i^2}$, $\psi_N$ is the normalized poloidal flux, which is 0 at the axis, $1$ at the edge, and $\psi_{\mathrm{ped}}$ at the pedestal top. Using an automated `fingerprints' \cite{Kotschenreuther2019} gyrokinetic mode identifier summarized in \Cref{tab:tab1}, we determine that the fastest growing mode at the pedestal top $\overline{\Delta} = 1/12$ (\Cref{fig:one}(a)) is a microtearing mode (MTM) \cite{Drake1980,Hardman2022}, typical of ELMy pedestals \cite{Dickinson2012}. In the pedestal half-width $\overline{\Delta} \in [1/4 - 3/4] $, \Cref{fig:one}(b) and (c) show the equilibrium is marginally unstable to KBMs \cite{Tang1980,Aleynikova2018}, with MTMs and impurity-gradient-driven ion-scale tearing modes dominating at lower $\alpha$ values. Strikingly, each flux surface in \Cref{fig:one}(a)-(c) is experimentally far below the ideal-ballooning stability boundary, evidence that the KBM clamps the pedestal profiles. KBM is a compelling candidate for limiting pressure profiles as it transports through both particle and energy channels, $D_s / \chi_s \sim 1$, where $D_s$ and $\chi_s$ are particle and heat turbulent diffusivities.

\begin{table}
\caption{\label{tab:example} Gyrokinetic instability classifications in this paper. $\mathcal{P}(A^{tb} _{\parallel }) = 1- |\int A_{\parallel} d\theta| / \int | A_{\parallel}| d\theta$ \cite{Hatch2012} and $\gamma$ is the linear growth rate. The `Tearing' mode is any ion-scale tearing-parity mode that is not an MTM \textcolor{black}{and $\beta = 8 \pi p / B^2$.}}
\begin{ruledtabular}
  \begin{tabular}{|| c | ccccc }
    \hline
    Mode & $\chi_i / \chi_e$ & $D_e / \chi_e$ & $D_i / \chi_i$ & $\mathcal{P}(A^{tb} _{\parallel })$ & $\partial \gamma / \partial \beta$ \\
    \hline
    KBM & $\sim1$ & $\sim1$ & $\sim1$ & $ 1$ & $>0$ \\ \hline
    MTM & $\ll 1$ & $\ll 1$  & $$ & $<1$ & $>0$ \\ \hline
    Tearing & $$ & $$ & $$ & $<1$ & $$ \\ \hline
    ETG & $\ll 1$ & $\ll 1$  & $$ & $ 1$ & $ <0$ \\
    \hline
  \end{tabular}
\end{ruledtabular}
\label{tab:tab1}
\end{table}

\textit{Equilibrium Variation.}-- While our radially \textit{local} analysis showed a NSTX discharge to be marginally KBM-unstable, we now describe the approach required to correctly find the steepest-possible pedestal profiles subject to kinetic instability. This approach combines pedestal width-height rescaling with self-consistent plasma equilibrium reconstruction. Following previous works \cite{Snyder2009}, we parameterize electron density and temperature profiles as
\begin{equation}
\begin{aligned}
& n_e(\psi_N) = n_{e,\mathrm{core}} \mathrm{H} \left[  \psi_{\mathrm{ped,n_e}}  -  \psi_N \right] (1-\psi_N^{\alpha_{n_1}})^{\alpha_{n_2}} + \\
& A_n \bigg{[} n_{e0} \left( t_2 - \tanh \left(\frac{\psi_N-\psi_{\mathrm{mid,n_e}}}{ S_{\Delta} \Delta_{n_e} /2} \right) \right) + n_{\mathrm{e,sep}} \bigg{]},
\label{eq:1}
\end{aligned}
\end{equation}
\begin{equation}
\begin{aligned}
& T_e(\psi_N) = T_{e,\mathrm{core}} \mathrm{H} \left[  \psi_{\mathrm{ped,T_e}} -  \psi_N \right] (1-\psi_N^{\alpha_{T_1}})^{\alpha_{T_2}} +  \\
& A_T \bigg{[} T_{e0} \left( t_2 - \tanh \left(\frac{\psi_N-\psi_{\mathrm{mid,T_e}}}{ S_{\Delta} \Delta_{T_e}/2} \right) \right) \bigg{]} + T_{\mathrm{e,sep}},
\label{eq:2}
\end{aligned}
\end{equation}
where $H$ is a Heaviside function, $n_{e,\mathrm{core}}$, $T_{e,\mathrm{core}}$, $n_{e0}$, and $T_{e0}$ are constants, $\Delta_{n_e}$ and $\Delta_{T_e}$ are the pedestal electron density and temperature widths -- which usually differ -- and ${\alpha_{\{n,T,J\},\{1,2\}}}$ are exponents. Quantities $A_{n}$, $A_{T}$, and $S_{\Delta}$ rescale the pedestal density, temperature, and width. The pedestal heights $n_{e,\mathrm{ped}}$ and $T_{e,\mathrm{ped}}$ are
\begin{equation}
n_{e,\mathrm{ped}} = n (\psi_{\mathrm{ped,n_e}}), \; T_{e,\mathrm{ped}} = T(\psi_{\mathrm{ped,T_e}}),
\end{equation}
$n_{e,\mathrm{sep}}$ and $T_{e,\mathrm{sep}}$ are evaluated at $\psi_N =1$, $t_2 = \tanh \left(2 \right)$, and $\psi_{\mathrm{ped},n_e} = \psi_{\mathrm{mid,n_e}} - \Delta_{n_e} / 2$. The current density $
J(\psi)$ is the sum of an Ohmic term $J_C (1-\psi_N^{\alpha_{J_1}})^{\alpha_{J_2}}$ with constants $J_C,{\alpha_{J_1}},{\alpha_{J_2}}$ and a bootstrap current term $J_{\mathrm{bs}}$ \cite{Sauter1999}.
 
\begin{figure}
        \centering
        \includegraphics[width=0.49\textwidth]{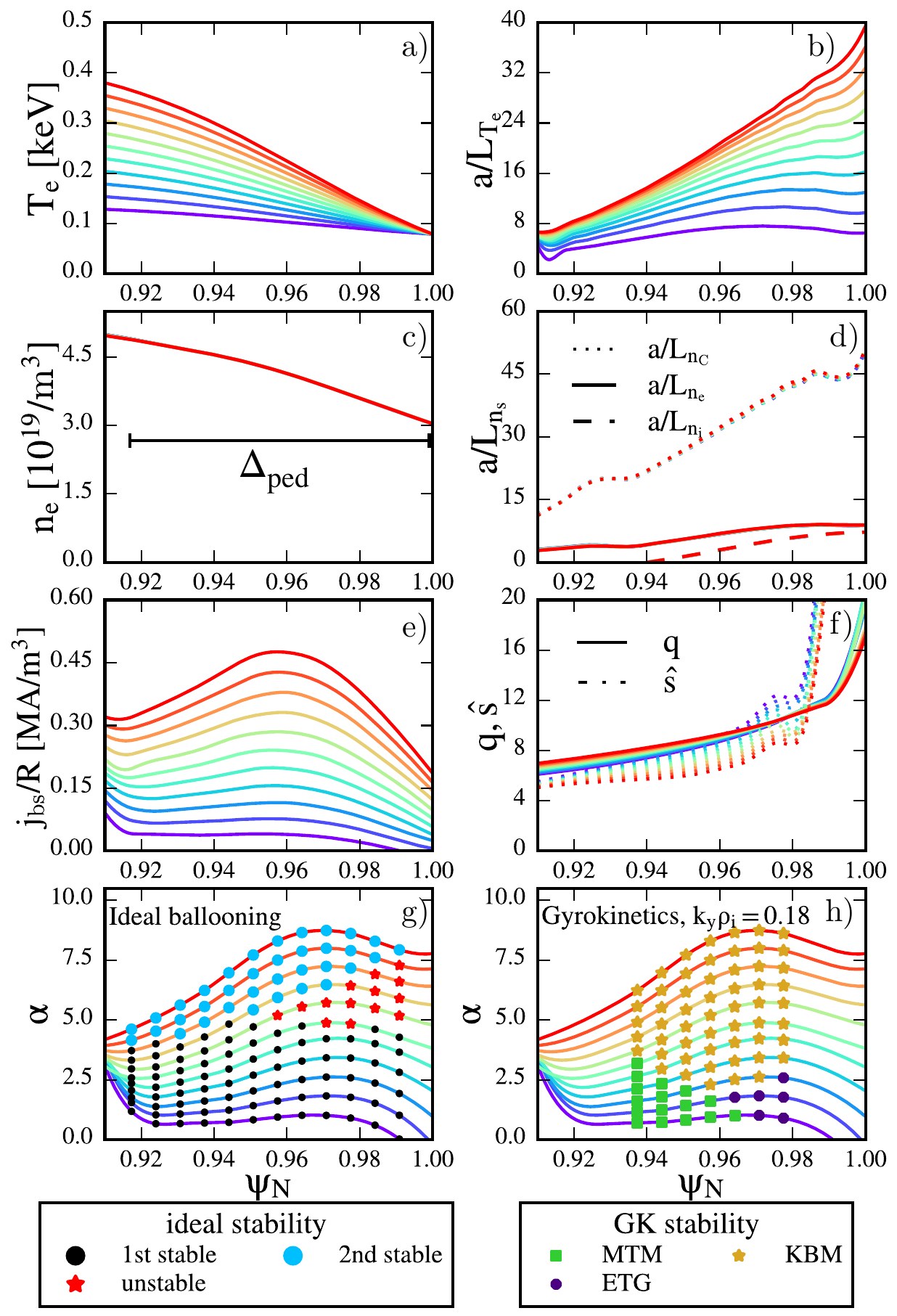}
        \caption{Quantities versus $\psi_N$ for equilibria with increasing $\overline{T}_{e,\mathrm{ped}}$ based on NSTX 139047 with fixed $\overline{n}_{e,\mathrm{ped}}$ and $\Delta_{\mathrm{ped}}$: a) $T_e$, b) $a/L_{T_e}$, c) $n_e$, d) $a/L_{n_s}$, e) $J_{\mathrm{bs}}/R$, f) safety factor $q$ and magnetic shear $s$, and $\alpha$ with ideal balooning stability (g)) and gyrokinetic stability for $k_y \rho_i  = 0.18$ (h)).}
        \label{fig:two}
\end{figure}

To distinguish between $n$ and $T$ contributions to pedestal pressure, we change the height with two bracketing cases of 1) varying $T$ at fixed $n$ and 2) varying $n$ at fixed $T$. Ion profiles are calculated using $\sum_s q_s n_s = 0$ and maintaining $T_i/T_e$. The rescaled pedestal heights $\overline{n}_{e,\mathrm{ped}}$ and $\overline{T}_{e,\mathrm{ped}}$ are obtained with coefficients $S_n$, $S_T$,
\begin{equation}
\overline{n}_{e,\mathrm{ped}} = S_{n} n_{e,\mathrm{ped}}, \;\;\;\;\; \overline{T}_{e,\mathrm{ped}} = S_{T} T_{e,\mathrm{ped}}.
\end{equation}
We choose width and height scalings $S_{\Delta} \in [0.4,1.6]$ and
\begin{equation}
\begin{aligned}
& \mathrm{fixed} \; T: \; S_n \in [0.4,1.6], \;\;  \mathrm{fixed} \; n: \; S_T \in [0.4,1.6].
\label{eq:4}
\end{aligned}
\end{equation}
While typically $\Delta_{n_e} \neq \Delta_{T_e}$, when rescaling the pedestal width, both $\Delta_{n_e}$ and $\Delta_{T_e}$ are rescaled by $S_{\Delta}$. Motivated by observations \cite{Leonard2017}, when rescaling the pedestal height, we fix $T_{e,\mathrm{sep}}$ and $n_{\mathrm{e,ped}}/n_{\mathrm{e,sep}}$. Equilibrium reconstruction is performed using EFIT-AI \cite{Lao2022}, where total plasma current $I_p$ and stored energy are conserved.

 \begin{figure}
    \centering
    \includegraphics[width=0.36\textwidth]{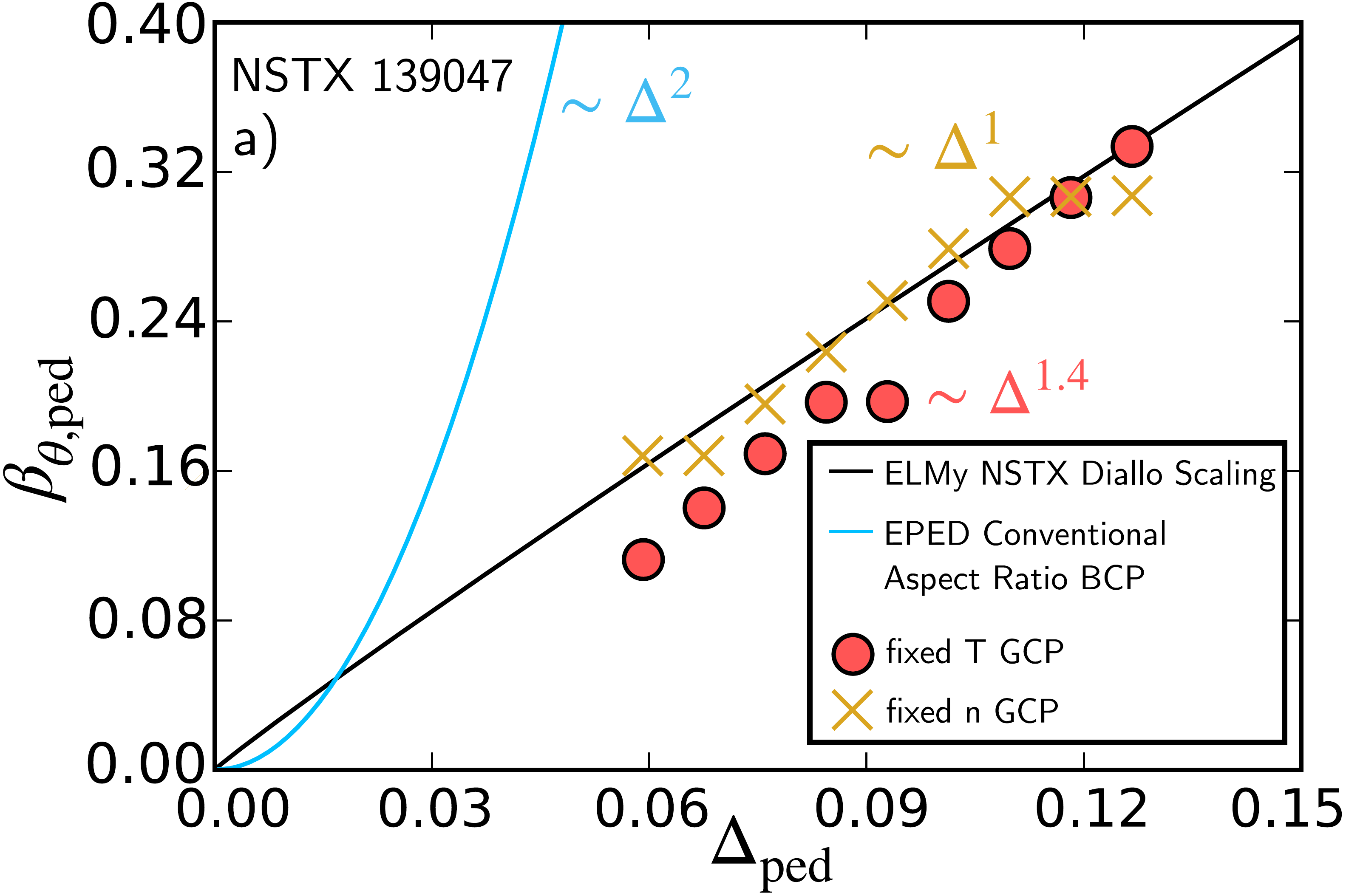}
    \begin{subfigure}[t]{0.5\textwidth}
        \centering
        \includegraphics[width=\textwidth]{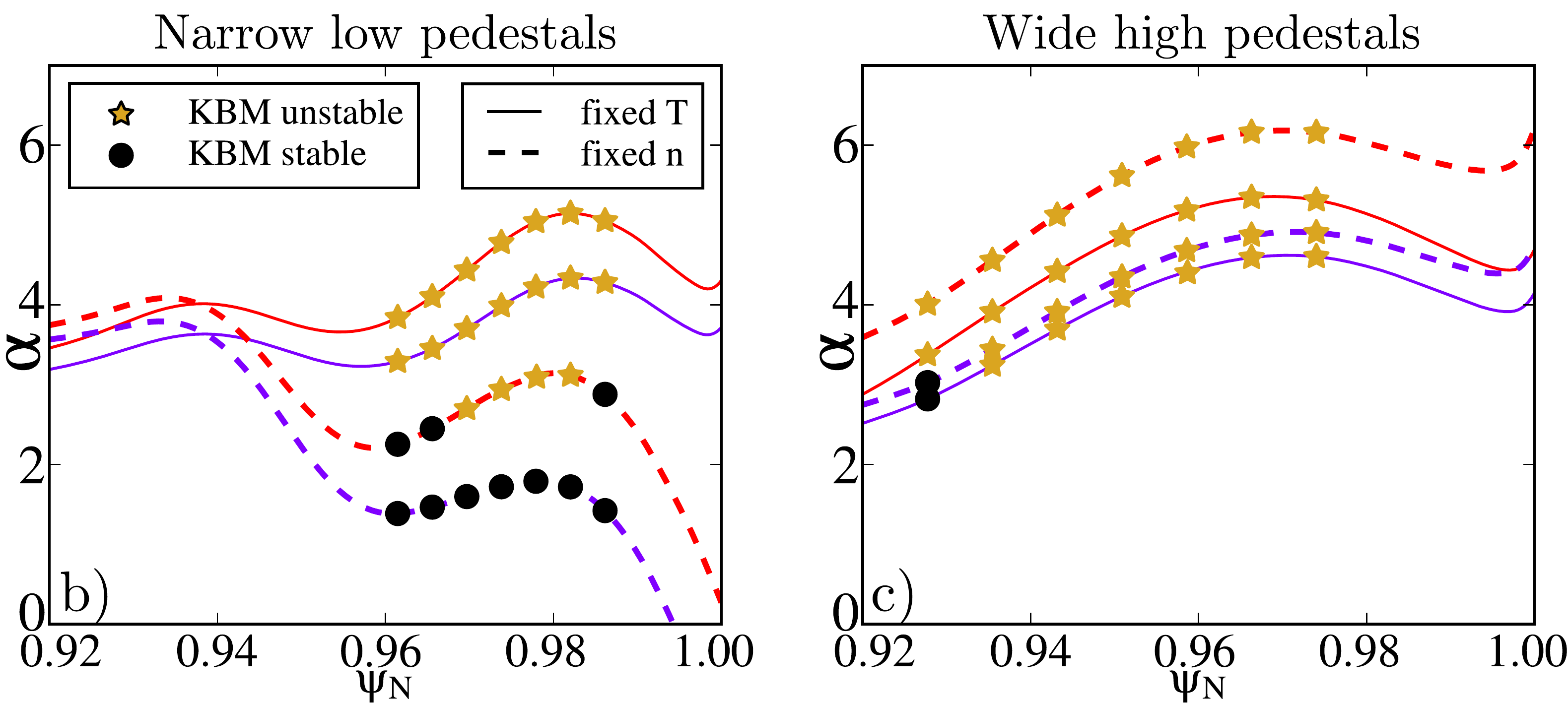}
    \end{subfigure}    
    \caption{ a): GCP equilibria for NSTX discharge 139047 for height variations with fixed $n$ (X markers) and fixed $T$ (circle markers), compared with the EPED and Diallo scalings. KBM stability and $\alpha$ across all $k_y \rho_i$ values for (b) narrow low pedestals and c) wide high pedestals.}
    \label{fig:three}
\end{figure}

Gyrokinetic flux-tube simulations are performed according to \Cref{eq:gke}, with wavenumbers where we found KBM to be most prevalent, $k_y \rho_i = 0.06,0.12,0.18$, radial wavenumber $k_x = 0$, using 3 kinetic species including electrons, and realistic up-down asymmetric geometry. The radial grid has 12 points, equally spaced in $\psi_N = [\psi_{\mathrm{mid}}-\Delta_{\mathrm{ped}}/2, \psi_{\mathrm{mid}}+ 5\Delta_{\mathrm{ped}}/12]$. The pedestal width and pedestal top location are $\Delta_{\mathrm{ped}} = (\Delta_{n_e} + \Delta_{T_e})/2, \;\;\; \psi_{\mathrm{ped}} = \psi_{\mathrm{mid}} - \Delta_{\mathrm{ped}} / 2$, where $\psi_{\mathrm{mid}} =  (\psi_{\mathrm{mid},n_e}+ \psi_{\mathrm{mid},T_e})/2$ and $\beta_{\theta, \mathrm{ped}}$, $p_{ped}$ are  $\beta_{\theta, \mathrm{ped}} = 8 \pi p_{\mathrm{ped}} /\overline{B}_{\mathrm{pol}}^2, \;\; p_{\mathrm{ped}} = 2 p_e (\psi_N = \psi_{\mathrm{ped}} )$. Here, $\overline{B}_{\mathrm{pol}} =4\pi I_p /  l c $ with flux surface circumference $l$.

\Cref{fig:two} shows equilibrium quantities for a height scaling of NSTX 139047 at fixed $\overline{ n}_{e,\mathrm{ped}}$ and fixed $\Delta_{\mathrm{ped}}$. \Cref{fig:two}(a) and (b) show increasing $T_e$ and $a/L_{T_s}$, while $n_e$ and $a/L_{n,s}$ in \Cref{fig:two}(c) and(d) are constant. Here, $1/L_{T_e} = -d \ln T_e / dr$. Steeper gradients generate a larger boostrap current, shown in \Cref{fig:two}(e), modifiying $q$ and reducing $s$, shown in \Cref{fig:two}(f). In \Cref{fig:two}(g) and (h), we plot gyrokinetic and ideal stability over $\alpha$ profiles. \Cref{fig:two}(g) shows larger $\alpha$ destabilizes the ideal-ballooning mode near the pedestal foot, while the pedestal center and top become second-stable. \Cref{fig:two}(h) shows the fastest growing gyrokinetic mode for $k_y \rho_i = 0.18$. Strikingly, the KBM can be the fastest growing gyrokinetic mode at roughly half the critical $\alpha$ for ideal instability. At lower $\alpha$ values, we find MTMs and electron-temperature-gradient (ETG) modes \cite{Dorland2000, Parisi2020, Adkins2022}.

\textit{The Gyrokinetic Critical Pedestal.}-- We now find the steepest pedestal profiles subject to gyrokinetic stability for NSTX discharge 139047, giving the relation between $\Delta_{\mathrm{ped}}$ and $\beta_{\theta, \mathrm{ped}}$. Analogous to the EPED Ballooning-Critical-Gradient (BCP) \cite{Snyder2009}, we hypothesise that the steepest pedestal profiles are marginally unstable to the same gyrokinetic instability across the pedestal half-width for any $k_y \rho_i \in [0.06, 0.12, 0.18]$, giving the Gyrokinetic Critical Pedestal (GCP). The instability should satisfy $D_s / \chi_s \sim 1$ \textcolor{black}{(although there are exceptions for marginal KBM \cite{Parisi2024stability})} and have similar linear and nonlinear critical thresholds \cite{Dimits2000}: KBM is a natural candidate \cite{Pueschel2010}, although a trapped-electron-mode \cite{Ernst2004, Hatch2015} may also be suitable. Our NSTX pedestal simulations found KBM dominates, so we here focus on KBM-limited profiles.

We find the GCP for the two bracketing cases of increasing pedestal height at fixed $n$ and $T$, shown in \Cref{fig:three}(a). While the fixed $n$ GCP profiles match NSTX measurements, the fixed $T$ GCP profiles under-predict $\Delta_{\mathrm{ped}}$ for lower $p_{\mathrm{ped}}$. This largely stems from our choice to fix $T_{e,\mathrm{sep}}$ while allowing $n_{e,\mathrm{sep}}$ to vary. For narrow low pedestals, fixing $T_{e,\mathrm{sep}}$ while varying  $\overline{T}_{e,\mathrm{ped}}$ for fixed $n$ height variation gives relatively lower $\alpha$ values than for fixed $T$ variation, shown in  \Cref{fig:three}(b). The higher $\alpha$ values for fixed $T$ make the KBM more unstable, hence giving a wider pedestal than for fixed $n$. For wide high pedestals, shown in \Cref{fig:three}(c), fixed $n$ and $T$ have similar $\alpha$ profiles and KBM stability properties near the GCP and hence predict similar pedestal profiles. There is also an effect due to bootstrap current modifying the magnetic shear, but since most NSTX pedestals we analyzed were in ideal first-stability, the change in the $\alpha$ profiles is likely more important.

\begin{figure}
        \centering
        \includegraphics[width=0.48\textwidth]{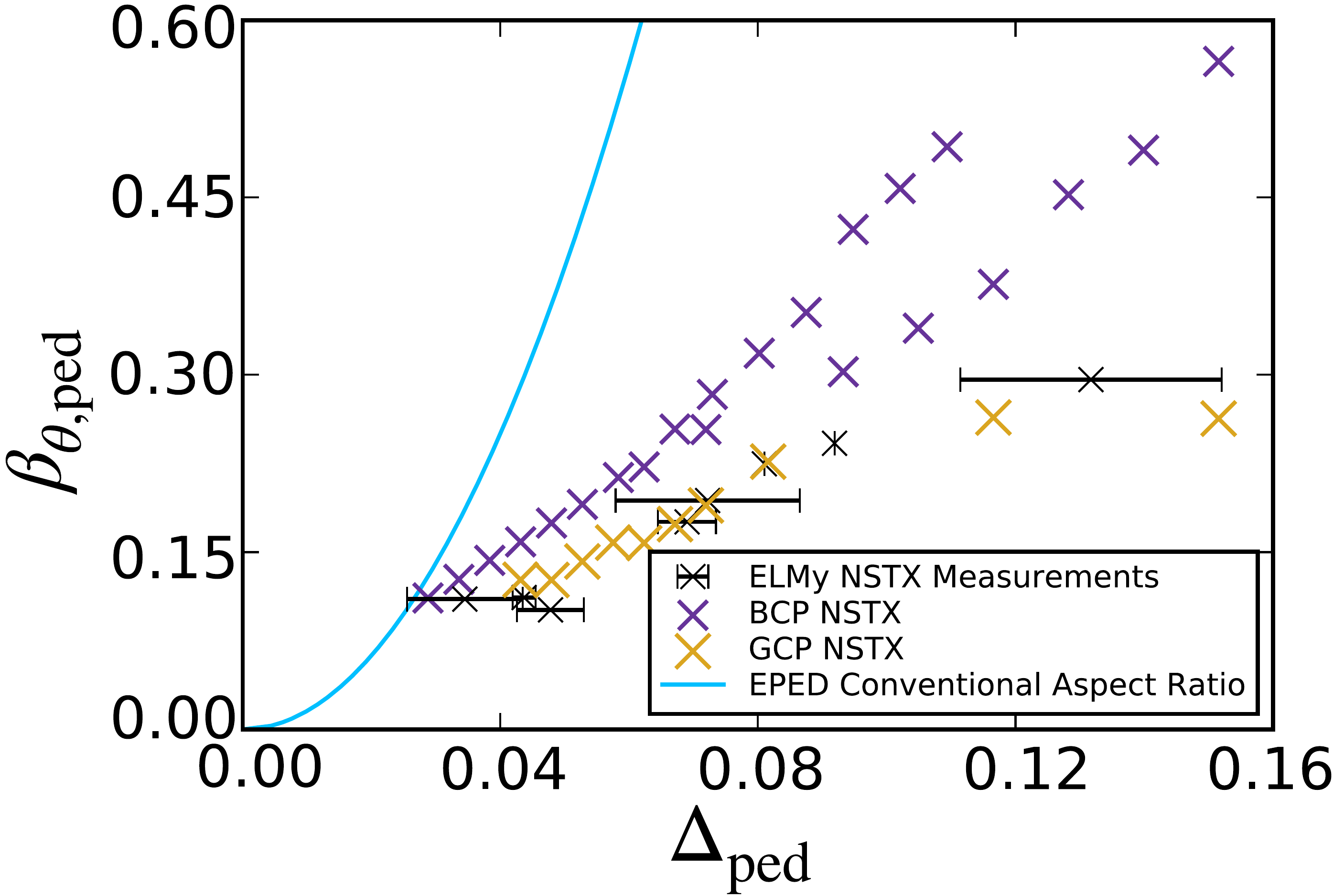}
        \caption{Width-height scaling for five NSTX discharges with fixed density variation. The GCP is in excellent agreement with ELMy NSTX measurements.}
        \label{fig:four}
\end{figure}

\textit{Averaged Pedestal Scalings.} -- We now find the GCP averaged over five NSTX discharges. In \Cref{fig:four}, we plot GCP equilibria points in gold calculated from NSTX discharges 130670, 132543, 139034, 139047, and 141300. Black markers indicate the Diallo scaling NSTX measurements $\Delta_{\mathrm{ped}} = (0.4\pm0.1) \beta_{\mathrm{ped}}^{1.05\pm0.2}$ \cite{Diallo2013}, in excellent agreement with our GCP scaling $\Delta_{\mathrm{ped}} \simeq 0.43 \beta_{\mathrm{ped}}^{1.03}$. We also perform ideal-ballooning stability calculations to obtain the BCP. The corresponding BCP -- purple markers in \Cref{fig:four} -- underpredicts pedestal width, $\Delta_{\mathrm{ped}} \simeq 0.26 \beta_{\mathrm{ped}}^{0.96}$. The dispersion of BCP datapoints is due to variations between NSTX discharges.

\begin{figure}
    \centering
     \includegraphics[width=0.5\textwidth]{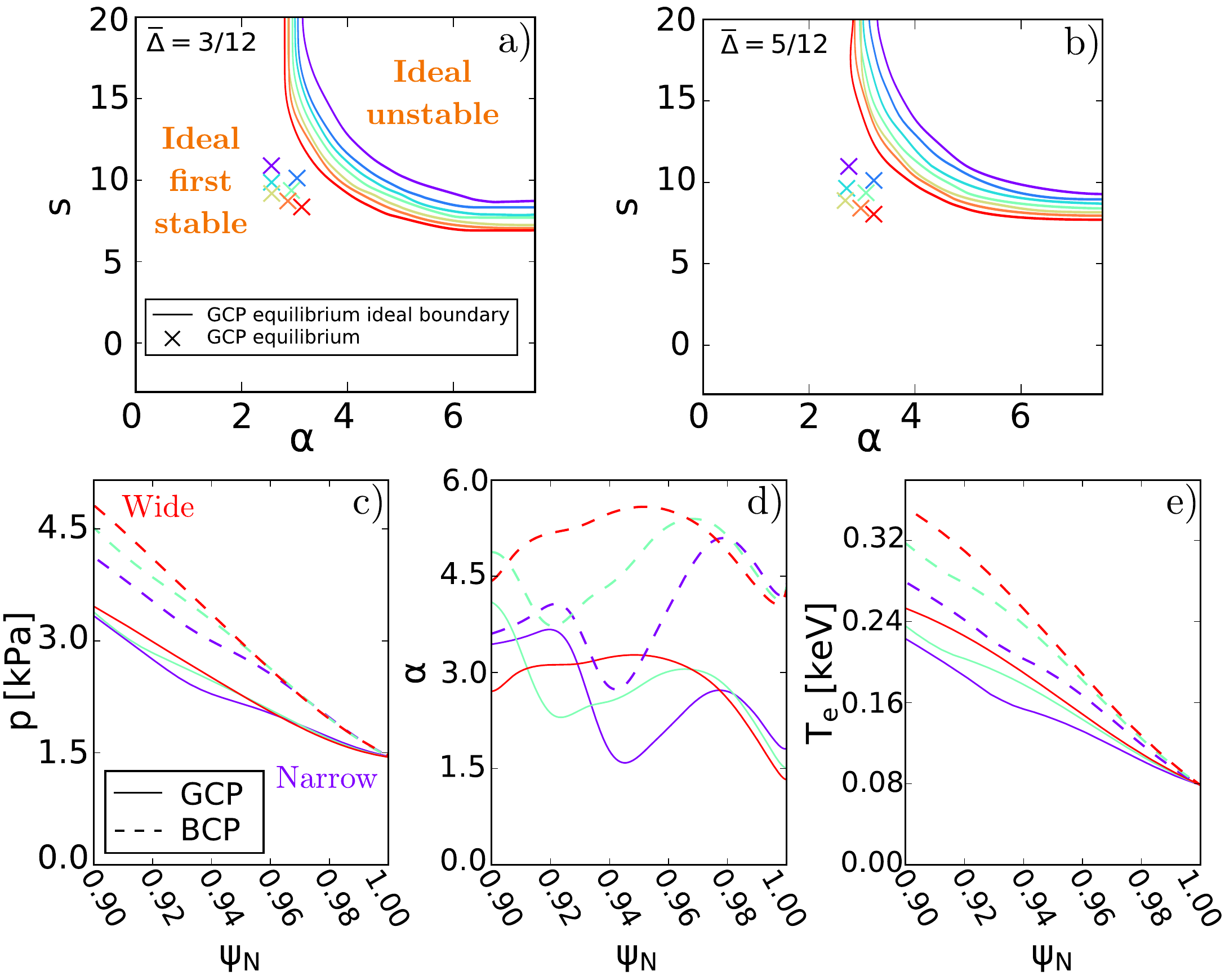}
    \caption{(a), (b): NSTX 139047 GCP equilibria in $s$-$\alpha$ coordinates for two radial locations: $\overline{\Delta} = 3/12, 5/12$. Markers indicate GCP equilibria and lines corresponding ideal stability boundaries. c), d), e): GCP and BCP pressure, $\alpha$, and $T_e$ profiles for three widths (by color), showing kinetic degradement of ideal-limited profiles.}
    \label{fig:five}
\end{figure}

KBM-limited pedestals are wider than ideal-ballooning-limited pedestals at fixed $\beta_{\theta,\mathrm{ped}}$ due to a lower $\alpha$ instability threshold for KBM. To elucidate this, in \Cref{fig:five} we show ideal $s-\alpha$ analysis for NSTX 139047 equilibria on the GCP. In \Cref{fig:five}(a) and (b), we plot the $s$ and $\alpha$ locations of GCP equilibria with different widths for two radial locations, $\overline{\Delta} = 3/12, 5/12$. The lines in \Cref{fig:five}(a) and (b) show the ideal stability boundaries for each equilibrium. The large distance in $\alpha$ between GCP equilibria and ideal-ballooning stability boundaries causes the large width prediction difference for the GCP and BCP in \Cref{fig:four}. In \Cref{fig:five}(c)-(e), we plot pressure, $\alpha$, and $T_e$ profiles for three GCP and three BCP equilbria with three different widths. The kinetic-ballooning-limited profiles can only support roughly one half of the pressure gradient than ideal-ballooning-limited profiles.

\begin{table}
\caption{\label{tab:example2} New pedestal scaling expressions and experimental measurements for NSTX, $\Delta_{\mathrm{ped}} = c_1 \left( \beta_{\theta, \mathrm{ped}} \right)^{c_2}$.}
\begin{ruledtabular}
  \begin{tabular}{|c||c|c|c|c|c| }
  & \shortstack{NSTX \\ Experiment \\ Diallo \cite{Diallo2013}} & \shortstack{Averaged \\ Fixed $n$ \\ GCP} & \shortstack{Averaged \\ Fixed $n$ \\ BCP}  & \shortstack{139047 \\ Fixed $n$ \\ GCP} & \shortstack{139047 \\ Fixed $T$ \\GCP} \\
    \hline
   \shortstack{$c_1$ \\ $c_2$} & \shortstack{$0.4\pm 0.1$ \\  $1.05 \pm 0.2$} & \shortstack{$0.43$ \\  $1.03$} & \shortstack{$0.26$ \\  $0.96$} & \shortstack{$0.39$ \\  $1.02$} & \shortstack{$0.27$ \\  $0.71$} \\ 
  \end{tabular}
\end{ruledtabular}
\label{tab:tab2}
\end{table}

\textit{Conclusion.}-- Combining self-consistent pedestal equilibrium variation with a gyrokinetic stability threshold model, we obtained a pedestal width-height scaling for kinetic-ballooning-limited NSTX spherical tokamak plasmas, $\Delta_{\mathrm{ped}} \simeq 0.43 \beta_{\theta, \mathrm{ped}}^{1.03}$, which is in excellent agreement with experimental measurements, $\Delta_{\mathrm{ped}} = (0.4\pm 0.1) \beta_{\theta, \mathrm{ped}}^{1.05 \pm 0.2}$ \cite{Diallo2013}. \textcolor{black}{Notably, using an ideal rather than a kinetic-ballooning threshold for NSTX underpredicts the width by 40\%, but the scaling $\Delta_{\mathrm{ped}} \simeq 0.26 \beta_{\theta, \mathrm{ped}}^{0.96}$ is still linear.} Our gyrokinetic model, applicable in the $\rho_{i}/a \to 0$ limit, is likely to become an even better approximation to future fusion reactors. While kinetic-ballooning-limited pedestals give lower average gradients than for ideal modes, their extra width might permit a larger $\beta_{\theta,\mathrm{ped}}$ before an ELM occurs \cite{Snyder2009,Osborne2015}. For sufficiently wide pedestals, there may even exist ELM-free regimes with other saturation mechanisms \cite{Chen2017,Maingi2017}, and thus paradoxically, as has been suggested in the context of other pedestal transport mechanisms \cite{Snyder2009, Osborne2015}, kinetic-ballooning-degraded pedestal profiles may increase $\beta_{\theta,\mathrm{ped}}$ and therefore fusion power. Recent results from ELM-free negative triangularity operation \cite{Austin2019} that aims to prevent ideal second stability access \cite{Nelson2022} may be modified by a larger KBM instability region \cite{Davies2022}, a phenomenon also seen in ST reactor studies \cite{Patel2022,Kennedy2023}. Scaling expressions presented in this paper are summarized in \Cref{tab:tab2}.

\textit{Code and data availability.} - Part of the data analysis was performed using the OMFIT integrated modeling framework \cite{OMFIT2015} using the Github projects \texttt{gk\_ped} \cite{Parisi2023a} and \texttt{ideal-ballooning-solver} \cite{Gaur2023a}. The data that support the findings of this study are openly available in Princeton Data Commons at https://doi.org/10.34770/vj4h-6120.

\textit{Acknowledgements.} - We are grateful for conversations with E. A. Belli, J. Candy, D. Dickinson, R. Maingi, J. E. Menard, M. J. Pueschel, P. B. Snyder, G. M. Staebler, and H. R. Wilson. We thank D. R. Hatch for a detailed reading of the manuscript. This work was supported by the U.S. Department of Energy under contract numbers DE-AC02-09CH11466, DE-SC0022270, DE-SC0022272, and the Department of Energy Early Career Research Program. The United States Government retains a non-exclusive, paid-up, irrevocable, world-wide license to publish or reproduce the published form of this manuscript, or allow others to do so, for United States Government purposes.

\bibliographystyle{apsrev4-1} %
\bibliography{EverythingPlasmaBib} %

\end{document}